\documentclass{jpp}
\usepackage{natbib}
\usepackage{graphicx}
\usepackage{color}
\usepackage{epstopdf, epsfig}
\usepackage{hyperref}
\usepackage{amsmath,amsfonts,amssymb}
\usepackage{subfigure}

\def\beq{\begin{equation}}
\def\eeq{\end{equation}}
\def\een#1{\label{#1} \end{equation}}
\def\beqa{\begin{eqnarray}}
\def\eeqa{\end{eqnarray}}
\def\ean#1{\label{#1} \end{eqnarray}}
\def\pd#1#2{\frac{\partial{#1}}{\partial{#2}}}
\def\od#1#2{\frac{d{#1}}{d{#2}}}
\def\eqref#1{(\ref{#1})}
\def\nnn{\nonumber \\}
\def\eps{\varepsilon}
\def\etal{{\emph{et al.}}}
\def\sech{{\rm sech}}

\shorttitle{Gardner equation and acoustic solitary waves}
\shortauthor{F. Verheest, W.\ A.\ Hereman}

\title{The Gardner equation and acoustic solitary waves in plasmas}

\author{
Frank Verheest\aff{1,}\aff{2}, 
Willy A.\ Hereman\aff{3}\corresp{\email{whereman@mines.edu}}
}

\affiliation{\aff{1}Sterrenkundig Observatorium, Universiteit Gent,
       Krijgslaan 281, B--9000 Gent, Belgium
\aff{2}School of Chemistry and Physics, University of KwaZulu-Natal,       
       Scottville, Pietermaritzburg 3209, South Africa
\aff{3}Department of Applied Mathematics and Statistics, Colorado School of Mines,
        Golden, Colorado 80401-1887, USA}

\begin{document}

\maketitle

\begin{abstract}
Ion-acoustic waves in a dusty plasma are investigated where it is assumed that  
the ions follow a Cairns distribution and the electrons are Boltzmann 
distributed. 
Two theoretical methods are applied: Sagdeev pseudopotential analysis (SPA) and 
reductive perturbation theory (RPT). 
Since SPA incorporates all nonlinearities of the model it is the most accurate 
but deriving soliton profiles requires numerical integration of Poisson's 
equation. 
By contrast, RPT is a perturbation method which at second order yields the 
Gardner equation incorporating both the quadratic nonlinearity of the KdV 
equation and the cubic nonlinearity of the modified KdV equation. 
For consistency with the perturbation scheme the coefficient of the quadratic 
term needs to be at least an order of magnitude smaller than the coefficient of 
the cubic term. 
Solving the Gardner equation yields an analytic expression of the soliton 
profile.
Selecting an appropriate set of compositional parameters, the soliton solutions 
obtained from SPA and RPT are analyzed and compared.  
\end{abstract}
%

%

\section{Introduction}
\label{intro}
%
For the theoretical treatment of electrostatic nonlinear solitary waves in 
plasmas there are essentially two methods: Sagdeev pseudopotential analysis 
(SPA) and reductive perturbation theory (RPT).
These methods predate their contemporary application in plasma physics in the 
mid 1960s.

SPA is commonly used in plasma physics to study the propagation of nonlinear 
solitary and periodic ion-acoustic waves.  
Based on an integration of the Poisson equation (which underlies all treatments 
of electrostatic plasma waves), one obtains a kind of energy integral, allowing 
a fully nonlinear analysis of one wave at a time.
The method draws on the analogy with classical mechanics, much as in the era of 
Newton, where the properties of the potential energy dictate the motion of a 
particle in a potential field.
The SPA method requires that the densities of the different plasma species can 
be expressed as functions of the electrostatic potential ($\varphi$) which is 
not always possible.

As the name suggests, RPT is a perturbation method which can be applied in many 
fields of the natural sciences, including fluid dynamics and plasma physics.
Based on RPT, solitary surface water waves were described by a Korteweg-de Vries 
(KdV) equation in 1895 \citep{KdV1895} to explain John Scott Russell's 
observations dating from 1834 \citep{ScottRussell1844}.
Once RPT was used (after a long period of dormancy) for solitary plasma waves 
in 1966 \citep{Washimi1966}, other nonlinear equations of KdV-type appeared such 
as the modified KdV (mKdV) \citep{Miura-etal-jmp-1968,Wadati-jpsjpn-1973} 
and Gardner \citep{Gardner1967,Gardner1974} equations. 
The Gardner equation is sometimes called a combined (or mixed) KdV-mKdV 
equation since both quadratic and cubic nonlinearities are present.
These equations, as well as the nonlinear Schr\"odinger and sine-Gordon 
equations, led to the discovery of elastically scattering waves (solitons) 
by \citet{Zabusky1965} and ingenious mathematical methods to compute them, most 
notably the inverse scattering transform discovered in 1967 \citep{Gardner1967}
(see, e.g. \citet{AblowitzClarkson1991}, \citet{AblowitzSegur1981}, and \citet{Remoissenet1999})  
and Hirota's method dating back to the early 1970s 
\citep{HirotaKdV1971,HirotamKdV1972,Hirota2004}.
The investigation of their rich mathematical structure revealed a whole range 
of properties similar to those of completely integrable dynamical systems.

Both the SPA and RPT methods have their advantages and drawbacks. 
When applicable, both methods can complement each other to give a fuller 
picture of the nature of the nonlinear wave solutions.
Of the two methods, SPA is the most accurate one because it uses the 
nonlinearities of the plasma model in full. 
One can still work with an analytical expression for the Sagdeev 
pseudopotential (the ``potential energy" in the mechanics analogy) but the 
profiles for the solitary waves have to be computed by numerical integration of 
Poisson's equation.

By contrast, RPT is entirely algorithmic and often leads to nonlinear evolution 
equations for which some properties and analytical profiles (for $\varphi$) are 
already known in the literature.
A drawback is that the nonlinearities are truncated to second or third order, 
making RPT less reliable to compute ion-acoustic waves with large amplitudes.

The purpose of this paper is twofold: 
(i) We will compare the results from SPA and RPT applied to a sufficiently 
complicated plasma model with compositional parameters such as mass, charge, 
and temperature.
Using SPA we will numerically compute soliton profiles for a suitable set of 
compositional parameters.  
Using RPT we will derive the Gardner equation. 
Its analytic soliton solutions will be compared with the numerical soliton 
profiles obtained from SPA for the same parameter values and for the same 
soliton velocity with respect to an inertial frame.
Although the literature about solitons computed with SPA and RPT 
{\textit{separately}} is vast, comparisons of the results from both treatments 
for the same plasma model are rare.
(ii) In the derivations of the Gardner equation, we will pay close attention to 
the choice of the compositional parameters which determines the signs and 
magnitudes of the coefficients of the quadratic and cubic terms. 
For the plasma model under consideration only the so-called {\textit{focusing}} 
Gardner equation is relevant. 
That is the equation that can be reduced to the focusing mKdV equation where 
the coefficients of all terms are positive (perhaps after scaling).
Consequently, for the compositional parameters used in this paper, ion-acoustic 
waves modeled by the Gardner equation can not take the shape of flat-top 
(sometimes called table-top) solutions \citep{HeremanGoktas-gardner-2024}. 
However, these table-top solitons arise in SPA as numerical solutions of  
Poisson's equation near double layers and triple root structures in some  
multispecies plasmas. 
The interested reader is referred to \citet{Verheest-etal-pp-2020} where 
table-top solutions of the model in this paper (and others) are discussed. 

The paper is organized as follows. 
Section~\ref{SPA-model} covers the governing equations of the plasma model 
under consideration.  
For an appropriate set of compositional parameters, SPA is used to numerically 
compute profiles of bright and dark solitons.   
The Gardner equation is derived using RPT in 
\S~\ref{RPT-gardner}. 
Close attention is paid to the magnitude of the coefficients of the quadratic 
and cubic terms in the Gardner equation to remain consistent with the terms 
retained within RPT. 
Using suitable compositional parameters, 
in \S~\ref{compare-SPA-RPT} 
the analytic soliton solutions of the Gardner equation are compared with the 
numerical soliton profiles based on SPA. 
Some conclusions are drawn 
in \S~\ref{conclusions}. 
In Appendix A we compare the results from applying SPA and RPT to a simple 
plasma model where the KdV equation (instead of the Gardner equation) is 
relevant. 

\section{Sagdeev pseudopotential analysis and plasma model}
\label{SPA-model}

We consider a dusty plasma \citep{VerheestASSL2000} consisting of cold charged 
negative dust, Boltzmann electrons and Cairns nonthermal ions. 
The model is written in normalized variables yielding a compact formulation 
where the relevant parameters are readily recognizable. 
In terms of the physics of the model, the normalized densities are really 
{\textit{charge}} densities but that has no impact on the mathematical analysis. 
The present model has been successfully used to study solitons in dusty plasmas 
\citep{SadhaNT12008a,SadhaNT22008b} and, more recently, for the correct 
description of nonlinear periodic (``cnoidal-like") waves in such plasmas 
\citep{CnoidalSadha2024}.
A similar approach can be readily applied to a wide range of other plasma 
models, where results from SPA and/or RPT are available, to establish their 
validity ranges.

Following the Cairns distribution \citep{Cairns1995}, at the macroscopic level 
the ion density $n_i$ is given by
\beq
n_i = (1 + \beta \varphi + \beta \varphi^2) \mathrm{e}^{-\varphi},
\label{ion}
\eeq
where $\varphi$ denotes the electrostatic potential and the nonnegative 
parameter $\beta$ measures the nonthermality.
Note that \eqref{ion} gives a deviation from the ubiquitous Boltzmann 
distribution which is included at the lower limit for $\beta=0$.
The very mobile electrons (with density $n_e$) are assumed to be Boltzmann 
distributed.
Thus, in normalized form, 
\beq
n_e = (1-f) \mathrm{e}^{\sigma \varphi},
\label{elec}
\eeq
where $\sigma = T_i/T_e$ is the ion-to-electron temperature ratio and $f$ is 
the fraction of the negative charge density taken up by the charged dust 
relative to the positively charged ions at equilibrium. 
Hence, $(1-f)$ represents the equilibrium electron charged density fraction.

Crucial in the analysis is the representation of the cold negative 
charged dust which, in a one-dimensional fluid description, comprises the 
equations of continuity,
\beq
\pd{n_d}{t} + \pd{}{x}(n_d u_d) = 0,
\label{cont}
\eeq
and momentum \citep{SadhaNT12008a}
\beq
\pd{u_d}{t} + u_d \pd{u_d}{x} = \pd{\varphi}{x},
\label{mom}
\eeq
where $n_d$ and $u_d$ are the density and velocity of the dust grains. 
The basic equations are coupled by Poisson's equation \citep{Watanabe1984}
\beq
\pd{^2\varphi}{x^2} + n_i - n_e - n_d = 0.
\label{poisA}
\eeq
To greatly simplify the mathematical analysis, we will work in a frame 
co-moving with the structure, by introducing 
\beq \zeta = x - V t, 
\label{zeta} 
\eeq 
where $V$ is the velocity of the nonlinear wave. 
In the Sagdeev formalism, it is assumed that solitary waves exist, with a 
stationary profile in the co-moving frame. 
For that, the restrictions on the parameters have to be established.

The time variable is subsumed in $\zeta$, and the description will only work if
the dust density $n_d$ can be expressed as function of $\varphi$.
Given that $n_e$ and $n_i$ are already of the appropriate form, once $n_d$ as 
function of $\varphi$ has been obtained, \eqref{poisA} becomes a differential 
equation from which $\varphi$ has to be determined.

We rewrite \eqref{cont} and \eqref{mom} in terms of $\zeta$ with the help of
\eqref{zeta} and integrate the resulting expressions with respect to $\zeta$, 
starting from the undisturbed equilibrium quantities faraway from the
structure. 
Hence, we impose the boundary conditions $\varphi = 0,$ $n_d = f,$ and 
$u_d = 0$ when $|\zeta| \to \infty$.
Eliminating $u_d$ between the two expressions thus obtained leads to 
\beq
n_d = \tfrac{f}{\sqrt{1+\tfrac{2 \varphi}{V^2}}},
\label{dust}
\eeq
involving a square root which is typical for cold plasma species.
    
Poisson's equation \eqref{poisA} then becomes
\beq
\od{^2\varphi}{\zeta^2} + 
(1 + \beta \varphi + \beta \varphi^2) \mathrm{e}^{-\varphi} 
- (1-f) \mathrm{e}^{\sigma \varphi}
- \tfrac{f}{\sqrt{1+\tfrac{2 \varphi}{V^2}}} = 0.
\label{pois}
\eeq
After multiplication by $d\varphi/d\zeta$ and integration with respect to
$\zeta$ one obtains an energy-like integral,
\beq
\frac{1}{2} \left( \od{\varphi}{\zeta} \right)^2 + S(\varphi) = 0, 
\label{energy}
\eeq
with the Sagdeev pseudopotential $S(\varphi)$ defined 
\citep{Sagdeev1966} as 
\beq
S(\varphi) =
1 + 3 \beta - (1 + 3 \beta + 3 \beta \varphi + \beta \varphi^2) \mathrm{e}^{-\varphi} 
  + \tfrac{1-f}{\sigma} \left( 1 - \mathrm{e}^{\sigma \varphi} \right) 
  + f V^2 \left( 1 - \sqrt{1+\tfrac{2 \varphi}{V^2}} \right).
\label{sagdA}
\eeq
Evidently, \eqref{pois} is then  
\beq
\od{^2\varphi}{\zeta^2} + S'(\varphi) = 0,
\label{poisagain}
\eeq
which plays a complementary role to \eqref{energy} in the investigation below.

The behavior of \eqref{sagdA} has to be studied as we vary the compositional 
parameters $f$, $\beta$ and $\sigma$.
One of the conditions to find soliton solutions is that the origin (at 
$\varphi=0$) is an unstable equilibrium, in other words, that $S(\varphi)$ is 
negative in the immediate neighborhood on the left and right of $\varphi=0$.
The conditions for that are $S(0)=0, S'(0)=0$ and $S''(0)<0$, where primes 
denote derivatives of $S$ with respect to $\varphi$.
$S(0)=0$ is obtained by adjusting the integration constants; 
$S'(0)=0$ follows from charge neutrality in equilibrium; 
and the concavity implied by $S''(0)<0$ requires that 
\beq
V^2 \geqslant V_a^2 = \frac{f}{1-\beta + (1-f)\sigma}.
\label{acous}
\eeq
This sets the acoustic velocity $V_a$ as the minimum soliton velocity, thus
solitary waves are superacoustic.  

One can easily check that $S(\varphi)\to -\infty$ for $\varphi\to +\infty$. 
Since $S(\varphi)<0$ near the origin, positive roots, if they exist, must occur 
in pairs.
When $V$ is sufficiently increased, the pair of positive roots closest to the 
origin becomes a double root.

In this model, $S(\varphi)$ does not have enough flexibility to have positive 
roots beyond that. 
Hence, the range of positive roots ends at the double root. 
In more complicated plasma models that is not necessarily the case but such 
scenarios are outside the scope of the present paper. 

We can also introduce a limit on the negative side,
\beq
\varphi_{\mathrm{lim}} = - \, \tfrac{1}{2} V^2,
\label{lim}
\eeq
obtained from \eqref{dust} at infinite dust compression ($n_d \to + \infty$).
To have a negative root of $S(\varphi)$ (before infinite dust compression 
occurs), one must have that $S(\varphi_{lim})\geqslant 0$, which yields another 
limit on possible values of $V$.   

To illustrate the shape of $S(\varphi)$ given in \eqref{sagdA}, we carefully 
select a set of compositional parameters,
\beq
\beta = 4/7, \qquad \sigma = 1/20, \qquad f = 0.61.
\label{parameters}
\eeq
For the nonthermality parameter $(\beta)$ there is an upper limit of $4/7$ 
in light of how the underlying microscopic Cairns distribution 
\citep{Cairns1995} has been defined. 
Selecting $\beta = 4/7$ produces a quite strong nonthermality.
Further details can be found in the corresponding soliton papers 
\citep{SadhaNT12008a,SadhaNT22008b}.
With respect to $\sigma = T_i/T_e$, one expects the heavier ions to have a 
lower temperature than the electrons which makes $\sigma = 1/20$ a reasonable 
choice. 

The choice of the third parameter ($f$) is motivated by our goal to compare the 
results from the application of SPA and RPT and the ensuing Gardner equation. 
As will be shown  
in \S~\ref{RPT-gardner}, 
respecting the conceptual 
constraints underlying the derivation of the Gardner equation, the coefficient 
$B$ of the quadratic nonlinearity should be close to zero whereas the 
coefficient $C$ of the cubic nonlinearity should be at least an order of 
magnitude larger than $B$. 
Specifically, $f$ has been selected so that the compositional parameters 
\eqref{parameters} produce $B\simeq 0.01$ and $C\simeq 0.5$. 

\begin{figure}
\centering
\includegraphics[width=42mm]{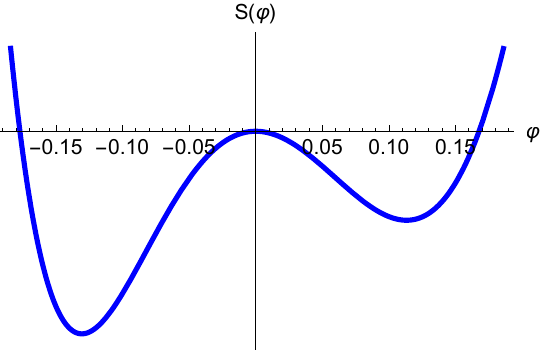} \;\; 
\includegraphics[width=42mm]{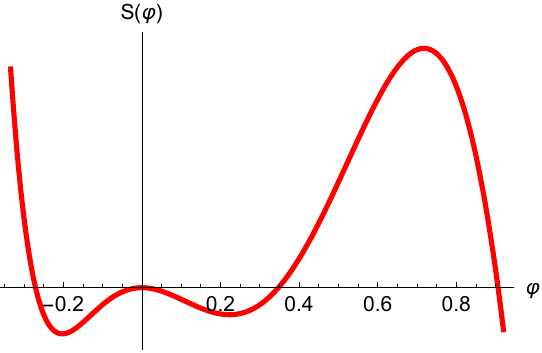} \;\;
\includegraphics[width=42mm]{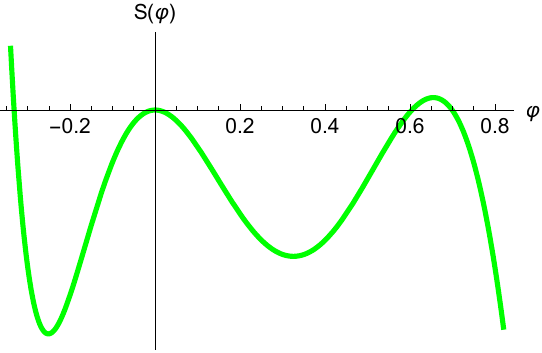}
\caption{
Graphs of the Sagdeev pseudopotential \eqref{sagdA} for $f = 0.61, \beta = 4/7, 
\sigma = 1/20$,   
and $V = 1.170$ (left), $V = 1.176$ (middle), $V = 1.182$ (right).
}
\label{SagdProfS}
\end{figure}

Continuing with SPA and inserting \eqref{parameters} into \eqref{acous} yields 
$V_a = 1.16679$.
Choosing then slightly larger values, namely $V = 1.170$, $V = 1.176$, and 
$V = 1.182$, enables us to plot the respective $S(\varphi)$ as shown in 
figure~\ref{SagdProfS}.  
%
We clearly see that there are positive roots, giving solitons with amplitudes 
$\varphi_{\mathrm{pos}} = 0.167704$, $0.347341$, $0.604500$, respectively.
These amplitudes are computed by numerically solving $S(\varphi) = 0$ with 
{\emph{Mathematica}}'s \verb|FindRoot| function. 
Using that same function, a numerical solution of $S(\varphi) = S'(\varphi) = 0$ 
for $V^2$ and $\varphi$ also shows that at $V_{\mathrm{dr}} = 1.18219$ 
a positive double root $\varphi_{\mathrm{dr}} = 0.6526$ is reached,
signalling the end of the range of solitons with positive amplitudes 
(the so-called ``bright" solitons).

Theoretically, there are either no or two positive roots, as discussed. 
So, the velocity range for bright solitons is $1.16679 \leqslant V < 1.18219$. 
For graphical clarity the larger of the two positive roots is not 
shown in the left graph in figure~\ref{SagdProfS} yet it exists, although 
without physical meaning, as it cannot be reached from the initial conditions 
at $\varphi=0$. 

There are also negative roots, giving rise to ``dark" solitons (with negative 
polarity for $\varphi$), 
$\varphi_{\mathrm{neg}}= -0.177029$, $-0.270600$, $-0.331836$, respectively, 
for the same compositional parameters. 
The range of negative roots is limited by the infinite dust compression, which 
is obtained from $S(\varphi_{\mathrm{lim}}) = S(-V^2/2) = 0$, yielding 
$V = V_{\mathrm{lim}} = 1.43927$ and, thus, $\varphi_{\mathrm{lim}} = -1.03575$.
Note that the dark solitons have larger amplitudes (in absolute value) and 
occur over a larger range for $V \geqslant V_a$, compared to the range for the 
bright solitons which disappears long before the range for dark solitons also 
ceases to exist.

\begin{figure}
\centering
\includegraphics[width=42mm]{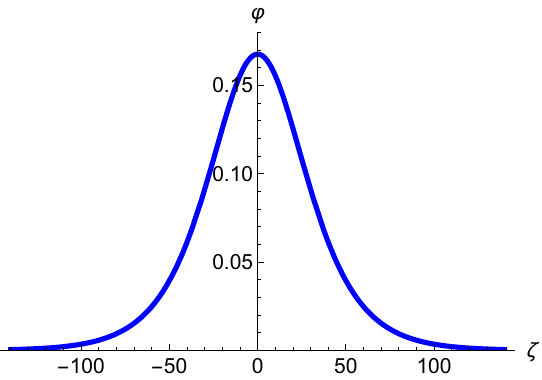} \;\;
\includegraphics[width=42mm]{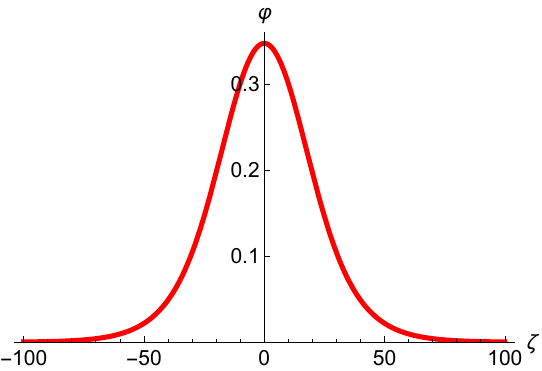} \;\;
\includegraphics[width=42mm]{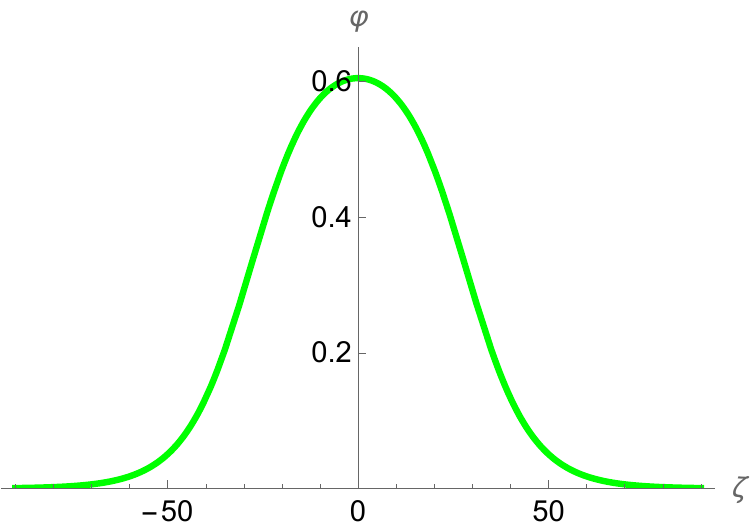}
\caption{
Graphs of bright solitons corresponding to the parameters given in 
figure~\ref{SagdProfS}.
}
\label{SagdPos}
\end{figure}

\begin{figure}
\centering
\includegraphics[width=42mm]{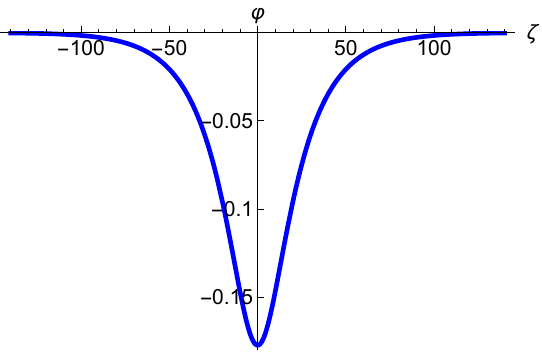} \;\;
\includegraphics[width=42mm]{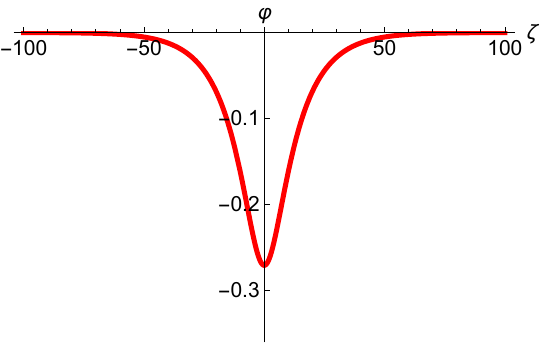} \;\;
\includegraphics[width=42mm]{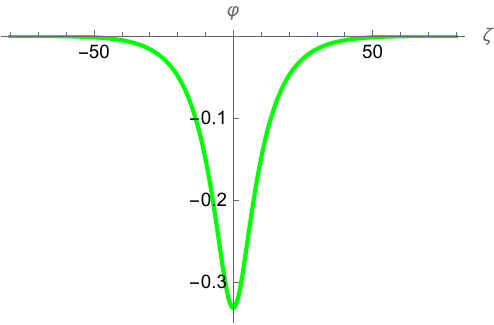}
\caption{
Graphs of dark solitons corresponding to the parameters given in 
figure~\ref{SagdProfS}.
}
\label{SagdNeg}
\end{figure}

The soliton profiles as shown in figures~\ref{SagdPos} and~\ref{SagdNeg} are 
based on numerical integration (with {\emph{Mathematica}}'s 
\verb|NDSolve| function) 
of Poisson's equation \eqref{pois} with conditions at the maxima or minima. 
In figure~\ref{SagdPos}, we used $\varphi^{\prime}(0) = 0$ together 
with $\varphi(0) = 0.167704$ (left), $\varphi(0) = 0.347341$ (middle), 
and $\varphi(0) = 0.604500$ (right).
In figure~\ref{SagdNeg}, we used $\varphi(0) = -0.177029$ (left), 
$\varphi(0) = -0.270600$ (middle), and $\varphi(0) = -0.331836$ (right), 
each again augmented with $\varphi^{\prime}(0) = 0$.
However, note that the scales in figures~\ref{SagdProfS},~\ref{SagdPos} 
and~\ref{SagdNeg} are different.
It is seen that the amplitudes of both the bright and dark solitons increase 
with $V$, but that the amplitudes of the bright solitons increase faster than 
those of the dark solitons (in the ranges where both polarities can be 
generated).  
%

For $V=1.17$, close to the acoustic speed $V_a$, the bright and dark solitons 
have more or less the same amplitudes (in absolute values). 
This is no longer true for larger $V$, where the bright solitons have larger 
amplitudes than the dark ones.
For $V = 1.182$ the bright soliton is wider and flatter. 
This becomes more and more noticeable as $V$ further increases 
and approaches $V_{\mathrm{dr}} = 1.182192261826$. 
Indeed, for $\varphi(0) = 0.6526$ and $V < V_{\mathrm{dr}}$ but very close 
to $V_{\mathrm{dr}}$, e.g. $V_{\mathrm{dr}} = 1.182192261825$ 
(i.e. $V_{\mathrm{dr}} - V = 10^{-12}$), the solution $\varphi(\zeta)$ is a 
wide flat-top soliton.
A discussion of these ``flatons" is outside the scope of this paper. 
The interested reader is referred to \citet{Verheest-etal-pp-2020}
for additional information. 

\section{Reductive perturbation theory and Gardner equation}
\label{RPT-gardner}
%
Application of RPT in plasma physics has lead to a host of nonlinear evolution 
equations of which three are prominent: the KdV equation itself, the mKdV 
equation with a cubic rather than quadratic nonlinearity, and the Gardner 
equation with both quadratic and cubic nonlinearities.
Each of these equations is completely integrable and exactly solvable with 
a panoply of methods.
Detailed studies of the structure, properties, and integrability of the KdV and 
mKdV equations \citep{AblowitzClarkson1991,Drazin1989}  
go back to the 1960s and the decades that follow 
\citep{Gardner1967,Gardner1974,Gesztesy1991}.
The Gardner equation is also completely integrable because it can be 
transformed into the mKdV equation with a Galilean transformation. 
Hence, a solution of the mKdV equation yields a solution of the Gardner equation, 
and vice versa. 
Using a generalized form of the Miura transformation (aka Gardner transformation), 
the Gardner equation can be transformed into the KdV equation, again confirming 
its complete integrability.  
However, that transformation will only be real-valued if the coefficient 
of the cubic term is positive (i.e. $C>0$ below). 
Furthermore, the Gardner transformation is non-reversible: 
from solutions of the Gardner equation one can obtain solutions of the
KdV equation, but not the other way around. 
A review of various integrability criteria, aforementioned transformations, 
and some solutions of the Gardner equation can be found 
in \citet{HeremanGoktas-gardner-2024}.
We refer the reader to \citet{Nasipuri-etal-ijp-2025} 
who give multi-soliton and breather solutions of a Gardner equation arising 
in an electron-positron-ion plasma model.  

Application of RPT for weakly nonlinear waves rests on two pillars:
First, a {\textit{stretching}} of the independent variables $x$ and $t$,
\beq
\xi = \eps (x - M t), \qquad \tau=\eps^3 t,
\label{stretch}
\eeq
where $M$ is a normalized velocity and $\eps$ is a bookkeeping parameter used 
to separate the orders of magnitude (i.e. smallness) of the various terms.  
Second, as with any perturbation method, {\textit{expansions}} of the dependent 
variables into smaller and smaller terms, 
\beqa
n_i &=& 1 + \eps n_{i1} + \eps^2 n_{i2} + \eps^3 n_{i3} + ...\ , \nnn
n_e &=& 1-f + \eps n_{e1} + \eps^2 n_{e2} + \eps^3 n_{e3} + ...\ , \nnn
n_d &=& f + \eps n_{d1} + \eps^2 n_{d2} + \eps^3 n_{d3} + ...\ , \nnn
u_d &=& \eps u_{d1} + \eps^2 u_{d2} + \eps^3 u_{d3} + ...\ , \nnn
\varphi &=& \eps \varphi_1 + \eps^2 \varphi_2 + \eps^3 \varphi_3 + ...\ ,
\label{expan}
\eeqa
where the ``constant" terms already have been inserted.
The expansions of $n_i$ and $n_e$ follow from the definitions of the ion and
electron densities in \eqref{ion} and \eqref{elec}, respectively, through the 
use of the expansion of $\varphi$ in \eqref{expan}.
Those for $n_d$ and $u_d$ need an interplay between \eqref{cont}, \eqref{mom}
and \eqref{poisA}.
Inserting the stretching yields
\beqa
&& \eps^3 \pd{n_d}{\tau} - \eps M \pd{n_d}{\xi} + \eps \pd{}{\xi}(n_d u_d) = 0, 
    \nnn
&& \eps^3 \pd{u_d}{\tau} - \eps M \pd{u_d}{\xi} + \eps u_d \pd{u_d}{\xi} 
    - \eps \pd{\varphi}{\xi} = 0, \nnn
&& \eps^2 \pd{^2\varphi}{\xi^2} + n_i - n_e - n_d = 0.
\label{basicD}
\eeqa
Substituting the expansions \eqref{expan} into the modified basic equations
\eqref{basicD} gives to second order the intermediate results
\beq
n_{d1} = -\, \frac{f \varphi_1}{M^2}, 
\qquad
u_{d1} = -\, \frac{\varphi_1}{M}. 
\label{nd1ud1}
\eeq
The integrations have been performed with zero boundary conditions for 
$\xi \to \pm\infty$, which are typical for solitons viewed in a co-moving 
frame, where the wave is centered at the origin and the wings vanish faraway on 
both sides.
These boundary conditions were also used  
in \S~\ref{SPA-model}. 
They are known as soliton boundary conditions and quite different from the 
conditions needed to generate nonlinear periodic waves 
\citep{OlivierVerheest-jpp-2022,CnoidalSadha2024}.
With \eqref{nd1ud1}, Poisson's equation \eqref{poisA} at order $\eps$ then 
yields the dispersion relation  
\beq
M^2 = M_a^2 = \frac{f}{1 - \beta + (1-f)\sigma}, 
\label{disp}
\eeq
fixing the wave speed in \eqref{stretch}.
Note that $M_a$ corresponds to the acoustic speed $V_a$ derived in \eqref{acous} 
in \S~\ref{SPA-model}, 
confirming the consistency between the two methods.
Rather than using the explicit expression \eqref{disp} for $M_a$ we will 
continue with the shorthand $M_a$ to keep the expressions more compact, in 
particular, those of the coefficients $A$, $B$, and $C$ given below.
At third order, the continuity and momentum equations yield 
\beq
n_{d2} = -\, \frac{f \varphi_2}{M_a^2} + \frac{3 f \varphi_1^2}{2 M_a^4}, 
\qquad
u_{d2} = -\, \frac{\varphi_2}{M_a} + \frac{\varphi_1^2}{2 M_a^3}. 
\label{nd2ud2}
\eeq
At order $\eps^2$ the Poisson equation then gives
\beq
\tfrac{1}{2} B \, \varphi_1^2 = 0,
\label{second}
\eeq
because the term in $\varphi_2$ vanishes by application of the dispersion law 
\eqref{disp}.
The constant  
\beq
B = 1 - (1-f)\sigma^2 - \frac{3 f}{M_a^4}
\label{B}
\eeq
in \eqref{second} is the coefficient of the quadratic nonlinearity which plays 
a crucial role in the distinction between the KdV, mKdV, and Gardner equations.
To make the term in \eqref{second} vanish, three possibilities should be 
considered: Either $B=0$, or $\varphi_1=0$, or $B$ is so small (i.e. order 
$\eps$) that the term in \eqref{second} should be included in Poisson's 
equation at order $\eps^3$.
We now discuss these scenarios in more detail. 

To continue with $\varphi_1\not= 0$ requires plasma models with enough 
parameters so that $B$ can be set to zero.
This can not be done, e.g. for ion-acoustic solitons in a simple plasma model 
where the ions are cold (no temperature effects) and the electrons are governed 
by a Boltzmann distribution (no inertial mass effects). 
An illustrative example is given in 
Appendix A. 
\subsection{The KdV equation}
If $B$ were nonzero (and finite), then the only possibility is to put 
$\varphi_1=0$ and recalibrate the description with $\varphi_2$ as the important 
variable. 
This would lead to the KdV equation, 
\beq
A \pd{\varphi_2}{\tau} + B \varphi_2\, \pd{\varphi_2}{\xi}
    + \pd{^3 \varphi_2}{\xi^3} = 0, 
\label{KdVo}
\eeq
describing a balance between slow time, nonlinear and dispersive effects. 
The compositional parameters are absorbed into coefficient $A$, given by
\beq
A = \frac{2 f}{M_a^3}, 
\label{A}
\eeq
and $B$ given in \eqref{B}.

Originally derived for solitons on the surface of shallow water by \citet{KdV1895}, 
the KdV equation appears in various physical contexts because it describes the 
propagation of nonlinear dispersive waves.
In particular, it has been used in plasma physics to model nonlinear 
ion-acoustic waves and solitons \citep{Varghese-etal-physd-2025},
resulting in a plethora of results in the 
literature for a great variety of multispecies plasmas.

Of course, by suitable scalings, for example,  
$X = \xi$, $T = \tau/A$, and $\varphi_2 = U/B$, 
\eqref{KdVo} can be replaced by 
\beq
\pd{U}{T} + U \pd{U}{X} + \pd{^3 U}{X^3} = 0, 
\label{KdVscaled}
\eeq 
with all coefficients equal to one and 
$U(X(\xi), T(\tau)) = B \varphi_2(\xi, \tau)$.
However, working with \eqref{KdVo} has the advantage that the coefficients are 
directly related to the compositional parameters which facilitates comparison 
with the plasma literature.
Regardless of the signs of $A$ and $B$, using the discrete symmetries 
$\tau \to -\tau$ and $\varphi_2 \to -\varphi_2$, \eqref{KdVo} can be 
transformed into the KdV equation where $A$ and $B$ are both positive.
See \citet{SinghKourakis2025} for a similar discussion of a slight variant of 
\eqref{KdVo}.
\subsection{The mKdV equation}
For certain plasma models, the parameters can be adjusted so that $B=0$,
requiring a different scaling and leading to the mKdV equation
\citep{Nakamura2009},
\beq
A \pd{\varphi_1}{\tau} + C \varphi_1^2 \pd{\varphi_1}{\xi}
    + \pd{^3 \varphi_1}{\xi^3} = 0,
\label{mKdV}
\eeq
having a cubic rather than a quadratic nonlinearity, with coefficient  
\beq
C = -\ \tfrac{1}{2} \left[1+3\beta+(1-f)\sigma^3 \right] + \frac{15 f}{2 M_a^6}.
\label{C}
\eeq
The change of variables $X = \xi$, $T = \tau/A$, and $\varphi_1 = U/\sqrt{|C|}$, 
transforms \eqref{mKdV} into 
\beq
\pd{U}{T} + \mathrm{sgn}(C) \, U^2 \pd{U}{X} + \pd{^3 U}{X^3} = 0,
\label{mKdVscaled}
\eeq
for $U(X(\xi), T(\tau)) = \sqrt{|C|} \, \varphi_1(\xi, \tau)$ and 
where $\mathrm{sgn}(C)$ denotes the sign of $C$. 
For $C>0$, \eqref{mKdVscaled} (and any scaled version of it) is called the 
{\textit{focusing}} mKdV equation which has soliton solutions of any order 
(see, e.g. \citet{AblowitzClarkson1991} and \citet{HeremanGoktas-gardner-2024}). 
The focusing mKdV equation has been extensively studied in plasma physics  
(see, e.g. \citet{VerheestHereman-jpp-2019} and \citet{Varghese-etal-physd-2025} 
and references therein).
If $C<0$, \eqref{mKdVscaled} is the {\textit{defocusing}} mKdV equation which, 
for example, describes the propagation of double layers or electrostatic shocks 
in plasmas \citep{Torven-prl-1981}.
The defocusing mKdV equation admits shock wave profiles (involving the 
$\mathrm{tanh}$-function) and table-top solutions 
(see \citet{HeremanGoktas-gardner-2024} and references therein). 
It is impossible to convert the defocusing mKdV equation into the focusing one 
by using discrete symmetries 
($\xi \to \pm\xi$, $\tau \to \pm\tau$, and $\varphi_1 \to \pm\varphi_1$, 
regardless of all possible combinations of signs). 

Both the KdV and focusing mKdV equations support waves that collide elastically 
(solitons), in principle for as many solitons as wanted
\citep{AblowitzClarkson1991,HeremanGoktas-hirota-2024}. 
A comparative study of ion acoustic waves in dusty plasma modeled by the KdV 
and mKdV-type equations can be found in 
\citet{KalitaDas2017} and \citet{Verheestetal-jpp-2016}. 
As an aside, replacing the quadratic and/or cubic terms with 
quartic and higher-order nonlinearities would destroy the complete 
integrability, and consequently, the typical soliton interactions would be lost 
\citep{Verheestetal-jpp-2016}.
\subsection{The Gardner equation}
We now turn our attention to the intermediate case where $B$ is not strictly 
zero but small enough so that quadratic as well as cubic nonlinearities are 
present and both play a significant role. 
This mixed (or combined) KdV and mKdV equation is often referred to as the 
Gardner equation \citep{Gardner1967,Gardner1974,Zabusky1965} which we will 
derive next. 

The momentum and continuity equations at fourth order yield  
\beqa
n_{d3} &=& -f \, \Big[ \frac{5 \varphi_1^3}{2 M_a^6} - \frac{3 \varphi_1 \varphi_2}{M_a^4} 
           + \frac{\varphi_3}{M_a^2} + \frac{2}{M_a^3} \int \pd{\varphi_1}{\tau}\, d\xi 
           \Big], \nnn
u_{d3} &=& - \frac{\varphi_1^3}{2 M_a^5} + \frac{\varphi_1 \varphi_2}{M_a^3} 
           - \frac{\varphi_3}{M_a} - \frac{1}{M_a^2} \int \pd{\varphi_1}{\tau} \, d\xi. 
\label{ud3nd3}
\eeqa
Substituting these expressions into Poisson's equation at order $\eps^3$,
one first encounters 
\beq
A \int \pd{\varphi_1}{\tau} \, d\xi + B \, \varphi_1 \varphi_2 
   + \tfrac{1}{3} C \, \varphi_1^3 + \pd{^2\varphi_1}{\xi^2} = 0
\label{prepreGardner}
\eeq
after setting $\varphi_3 = 0$.
The coefficients $A, B,$ and $C$ in \eqref{prepreGardner} are given in \eqref{A},
\eqref{B}, and \eqref{C}, respectively. 
Given the smallness of $B$ (close to the critical case $B=0$ leading to the 
mKdV equation \eqref{mKdV}) the term $B \varphi_1 \varphi_2$ is of higher order 
and should be discarded.
The same argument holds for the term in \eqref{second}, which should have 
been ``upgraded" to the next higher order and therefore be included. 
Hence, \eqref{prepreGardner} should be replaced by
\beq
A \int \pd{\varphi_1}{\tau} \, d\xi + \tfrac{1}{2} B \, \varphi_1^2  
   + \tfrac{1}{3} C \, \varphi_1^3 + \pd{^2\varphi_1}{\xi^2} = 0,
\label{preGardner} 
\eeq
which, after differentiation with respect to $\xi$, yields the true Gardner 
equation,
\beq
A \pd{\varphi_1}{\tau} + B \varphi_1 \pd{\varphi_1}{\xi} 
+ C \varphi_1^2 \pd{\varphi_1}{\xi} + \pd{^3\varphi_1}{\xi^3} = 0, 
\label{Gardner} 
\eeq
where for consistency, $B$ should be small (i.e. same order as $\varphi_1$) 
and $C$ should be of order unity.
If in \eqref{Gardner} $B$ and $C$ were both finite (viz.\ order unity) the 
quadratic term with coefficient $B$ would dominate and the cubic term with 
coefficient $C$ would be a negligible correction!   
Thus, for consistency, the Gardner equation requires that $|B \varphi|$ is of 
the same order of smallness as $|C \varphi^2|$.
If not, one of the two terms dominates and that would yield solutions 
reminiscent of the KdV or mKdV solitons.
This is of particular importance when the Gardner equation models a physical 
process where the higher-order nonlinearities have been neglected. 

Without loss of generality, we continue with $B > 0$ in (\ref{Gardner}) 
because, if $B < 0$, replacing $\varphi_1$ by $-\varphi_1$ would make the 
coefficient of $\varphi_1 \pd{\varphi_1}{\xi}$ positive again. 
The change of variables 
$X = B \xi /\sqrt{|C|}$, $T = B^3 \tau/(A |C| \sqrt{|C|})$, and 
$\varphi_1 = B U/|C|$, allows one to replace \eqref{Gardner} by 
\beq
\pd{U}{T} + U \pd{U}{X} + \mathrm{sgn}(C)\, U^2 \pd{U}{X} + \pd{^3 U}{X^3} = 0, 
\label{Gardnerscaled}
\eeq
for $U(X(\xi), T(\tau)) = (|C|/B) \, \varphi_1(\xi, \tau)$.
In analogy to the mKdV equation, \eqref{mKdVscaled} is called focusing or 
defocusing depending on whether the sign of $C$ is positive or negative. 
No discrete symmetries of $\xi, \tau,$ or $\varphi_1$ will flip the sign of the 
coefficient of $\varphi_1^2 \pd{\varphi_1}{\xi}$. 
So, the cases $C>0$ and $C<0$ would have to be treated separately. 

The Gardner equation has many applications  
\citep{HeremanGoktas-gardner-2024,Zhang-etal-rmp-2014} 
ranging from fluid dynamics to plasma physics \citep{OlivierVerheest-pp-2020}.
In the study of double layers and near-critical plasma compositions the 
defocusing Gardner equation plays a role \citep{Olivier-etal-jpp-2016}. 
For the plasma model treated in this paper and variants thereof only the 
focusing Gardner equation is relevant 
\citep{BachaTribeche-jpp-2013,Gilletal-jpp-2005,XieHe-pp-1999}.

\section{Comparison of the results from SPA and RPT}
\label{compare-SPA-RPT}

After having examined both methods from a theoretical point of view 
in the previous two sections, we are now ready to numerically compare 
the results obtained from SPA with these from RPT. 
Although we restrict our comparison to the model at hand, our approach is 
applicable to other multispecies plasma models with a sufficient number of 
compositional parameters.

Recall that RPT requires that $M = M_a$ with $M_a$ defined in \eqref{disp}.
Thus, in \eqref{stretch}, $M_a$ is the linear wave speed with respect to the 
laboratory inertial frame for the ``space" variable ($\xi$). 
Hence, the velocity $v$ of soliton solutions of \eqref{Gardner} is measured 
with respect to that frame.
By contrast, in SPA the soliton speed $V$ refers to the inertial laboratory 
frame as defined in \eqref{zeta}.
Regardless of the definition, for acoustic wave modes the soliton speed is 
always superacoustic (that is, larger than the original acoustic velocity).

Using model parameters \eqref{parameters}, we compute \eqref{B}, \eqref{A}, and 
\eqref{C} and insert these into \eqref{Gardner} yielding  
\beq
0.768044\, \pd{\varphi_1}{\tau} + 0.0116414\, \varphi_1 \pd{\varphi_1}{\xi}
+ 0.456023\, \varphi_1^2 \pd{\varphi_1}{\xi} + \pd{^3\varphi_1}{\xi^3} = 0.
\label{modGard}
\eeq
The analysis that follows is based on the well-known solitary wave solution of 
the Gardner equation \eqref{Gardner} in the form 
(see, e.g. \citet{HeremanGoktas-gardner-2024} and 
\citet{Olivier-etal-jpp-2016})
\beq
\varphi_1(\xi,\tau)
  = \frac{6 k^2}{B[1 + \sqrt{1+\frac{6 C}{B^2}k^2} \cosh(k(\xi- \tfrac{k^2}{A} \tau))]}  
  = \frac{6 A v}{B[1 + \sqrt{1+\frac{6 A C}{B^2} v} \cosh(\sqrt{A v}(\xi - v \tau))]}, 
\label{solGardner}
\eeq
since the wave number $(k)$ and wave speed $(v)$ are linked by $v=k^2/A$.

To compare the graphs of the solutions of the Gardner equation with those 
based on Sagdeev's approach, as noted above, the velocities refer to 
different moving frames, that is, $V = V_a + v$. Hence, $v = V - V_a$ 
which we will use below.  
As mentioned below \eqref{disp}, $M = M_a = V_a$. 
So, with regard to \eqref{zeta} and \eqref{stretch}, 
$\zeta = x - V t$ $ = x - V_a t - v t$ $ = \xi - v \tau $, after setting 
the bookkeeping parameter $\eps$ equal to $1$. 
When the values for $A, B$ and $C$ are inserted in \eqref{solGardner} the 
soliton profiles can be plotted. 

\begin{figure}
\centering
\includegraphics[width=42mm]{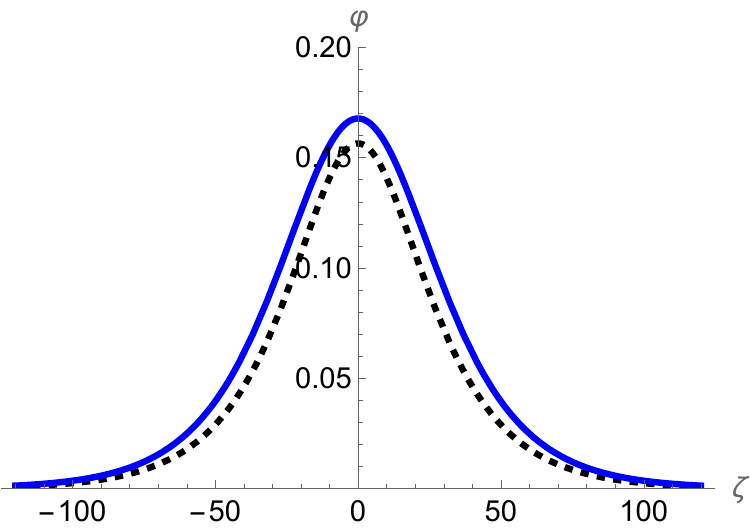} \;\;
\includegraphics[width=42mm]{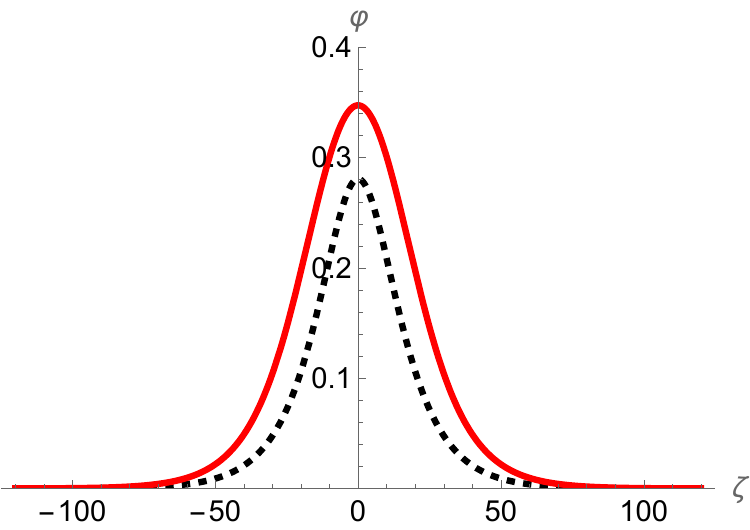} \;\;
\includegraphics[width=42mm]{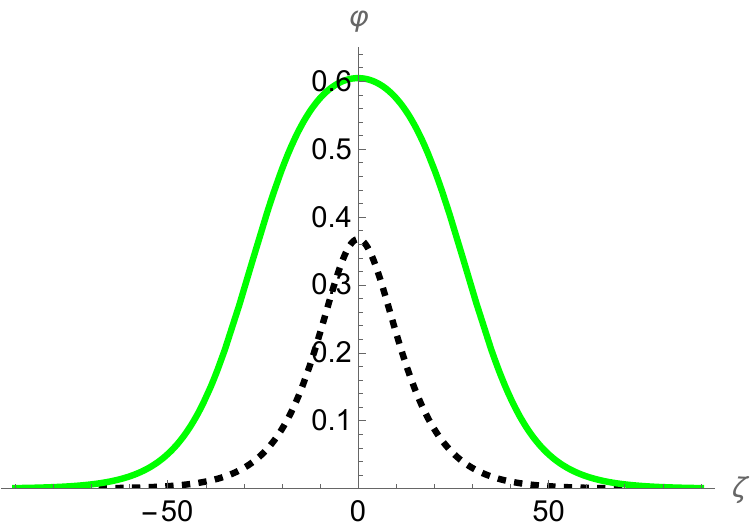}
\caption{
Graphs of bright solitons for the parameters given in figure~\ref{SagdProfS} 
but computed with two different techniques: 
Sagdeev's pseudopotential approach yields the solid curves 
(copied from figure~\ref{SagdPos}) 
and the solution \eqref{solGardner} of Gardner's equation gives the dashed curves, 
using $v = 0.00321$ (left), $v = 0.00921$ (middle), and $v = 0.01521$ (right).
}
\label{combineSG}
\end{figure}

In figure~\ref{combineSG}, we have combined the graphs obtained by 
SPA and RPT using $\zeta = \xi - v \tau$ as a single argument.
Recall that $V_a = 1.16679$.  
Hence, to compare with the graphs in figure~\ref{SagdPos}, we must 
evaluate \eqref{solGardner} for 
$v = 1.170 - 1.16679 = 0.00321$, 
$v = 1.176 - 1.16679 = 0.00921$, 
and $v = 1.182 - 1.16679 = 0.01521$.
It is seen that for larger $V$ and corresponding $v$, the solitons obtained 
with SPA are taller and much wider than those derived from Gardner's equation 
but both have the usual property that increasing amplitudes (corresponding to 
increasing velocities) result in reduced widths.
As $V$ gets closer and closer to $V_{\mathrm{dr}} = 1.18219$ the solutions 
of Gardner's equation deviate more and more from the solitons obtained from SPA
with amplitudes approaching $\varphi_{\mathrm{dr}} = 0.6526$.
From these comparisons one might conclude that the solutions of Gardner's equation 
are quite reliable up to $\varphi \simeq 0.2$. 

Unfortunately, we have been unable to compute dark soliton solutions of 
Gardner's equation that vanish at $\pm \infty$.
Hence, a correspondence with the dark solitons based on Sagdeev's approach can 
not be established. 

\section{Conclusions}
\label{conclusions}
%
In this paper we have investigated ion-acoustic waves in a dusty plasma with 
Cairns distributed ions and Boltzmann distributed electrons. 
We have applied Sagdeev pseudopotential analysis (SPA) and reductive 
perturbation theory (RPT). 
The SPA method 
retains all nonlinearities of the model and therefore yields the most 
accurate results but requires a numerical integration of Poisson's equation to 
get soliton profiles. 
By contrast, the accuracy of the results from RPT depends on the order 
of nonlinearity taken into account. 
The larger the number of terms retained in the perturbation expansions the more 
accurate the results will be but the harder it becomes to find analytic 
solutions along the way.  
Keeping terms up to second order, RPT yields the Gardner equation 
\eqref{Gardner} which still can be solved analytically and therefore yields a 
closed-form expression of the soliton profile. 

The derivation of the Gardner equation must be done with care. 
First, the plasma model must have a sufficient number of compositional 
parameters for the Gardner equation to be applicable. 
Second, we have shown that for consistency with the perturbation treatment, the 
coefficient ($B$) of the quadratic term should be at least an order of 
magnitude smaller than the coefficient ($C$) of the cubic term. 
If $C$ is of order unity and $B$ were of the same order, the quadratic term 
would prevail over the cubic term, which could then be neglected (leading to 
the KdV equation).  
Here again, a multispecies plasma should have enough compositional parameters 
to allow for a tiny $B$ and a positive $C$. 
Given the plethora of multispecies plasma models available in the literature 
(see, e.g. the references in \citet{Nasipuri-etal-ijp-2025}),   
there certainly are models that satisfy this requirement.

For an appropriate set of compositional parameters, the solitons obtained 
with SPA and RPT have been analyzed and compared. 
Although such comparisons are rarely done in the literature, they reveal 
important information about the range of validity of the commonly-used 
soliton solution of the Gardner equation. 
For the model in this paper, the discrepancies between the two methods indicate 
that the Gardner soliton is of limited use at higher amplitudes. 
We expect this also to be true in various other multispecies plasma models 
where the Gardner is derived via RPT. 
Careful investigation of the signs of the coefficients in the equation 
and estimation of their magnitudes are warranted.
A comparison of the results from SPA and RPT is also recommended because it  
will provide additional insight in the usefulness of analytic solutions.     

In Appendix A it is shown that simple ion-acoustic plasma models do not have 
enough compositional flexibility to go beyond the KdV equation. 
Based on our investigation we conclude that in simple plasma models the KdV or 
mKdV equations are the relevant ones, not the Gardner equation.  
 
   
\vfill 
\newpage
\section*{Acknowledgements}
%
FV thanks North West University (Department of Mathematics, Mahikeng Campus) 
and his host C.\ P.\ Olivier, and South African National Space Agency 
(Hermanus Space Science Campus) and his host S.\ K.\ Maharaj for their 
kind and warm hospitality, during a stay where part of this work was discussed.
We are grateful to the anonymous reviewers for their positive comments and
valuable suggestions.

{\it Editor Thierry Passot thanks the referees for their advice in evaluating 
this article}.

\section*{Funding}
This research received no specific grant from any funding agency, commercial or
not-for-profit sectors.
\section*{Declaration of interests}
The authors report no conflict of interest.
\section*{Author contributions}
Both the authors contributed equally to the analysis, reaching conclusions and 
in writing the paper.
%

\section*{Appendix A. Simple ion-acoustic waves}
\renewcommand{\theequation}{A.\arabic{equation}}
\setcounter{equation}{0} 
In this appendix we use the simplest model of ion-acoustic solitons in a plasma 
consisting of electrons with Boltzmann distribution,
$n_e = \mathrm{e}^{\varphi}$, and cold ions.
The model is then governed by the ion equations expressing continuity,
\beq
\pd{n_i}{t} + \pd{}{x}(n_i u_i) = 0,
\label{contI}
\eeq
and momentum,
\beq
\pd{u_i}{t} + u_i \pd{u_i}{x} + \pd{\varphi}{x} = 0,
\label{momI}
\eeq
coupled by Poisson's equation
\beq
\pd{^2\varphi}{x^2} + n_i - \mathrm{e}^{\varphi} = 0.
\label{poisI}
\eeq
These equations also follow from the dusty plasma model discussed in the 
preceding sections by putting $f=1$ (so that $\sigma$ disappears), $\beta=0$, 
and interchanging the polarity of the charged particles ($\varphi \to -\varphi$).
In this simplest model for ion-acoustic solitons there are no compositional 
parameters to select since all have ``disappeared" in the normalization.
\subsection*{Sagdeev pseudopotential analysis}
\begin{figure}
\centering
\includegraphics[width=42mm]{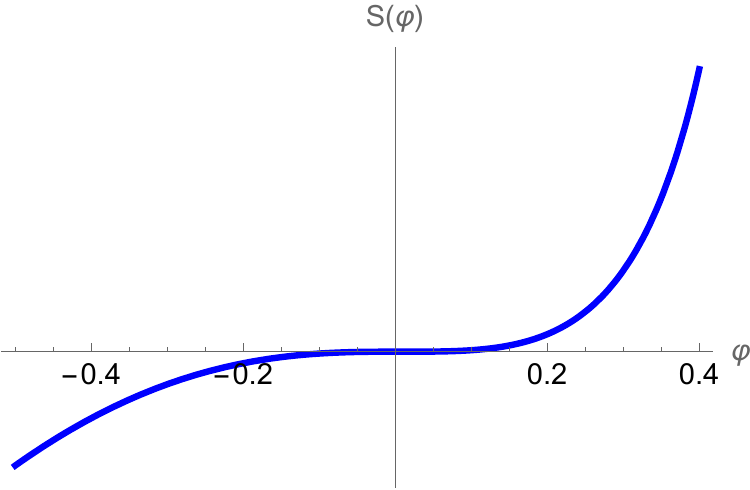} \;\;
\includegraphics[width=42mm]{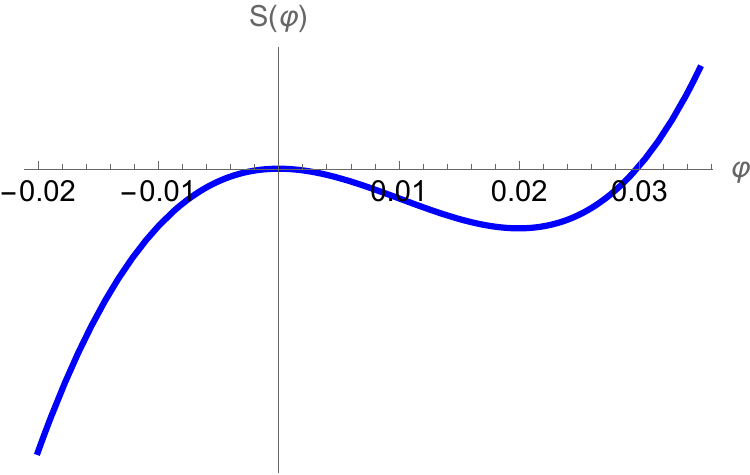} 
\caption{
Graph of the Sagdeev pseudopotential \eqref{sagdI} for $V = 1.01$ (left) 
and a zoom near the root $\varphi = 0.02978$ (right).
}
\label{SagdKdVd1}
\end{figure}

\begin{figure}
\centering
\includegraphics[width=42mm]{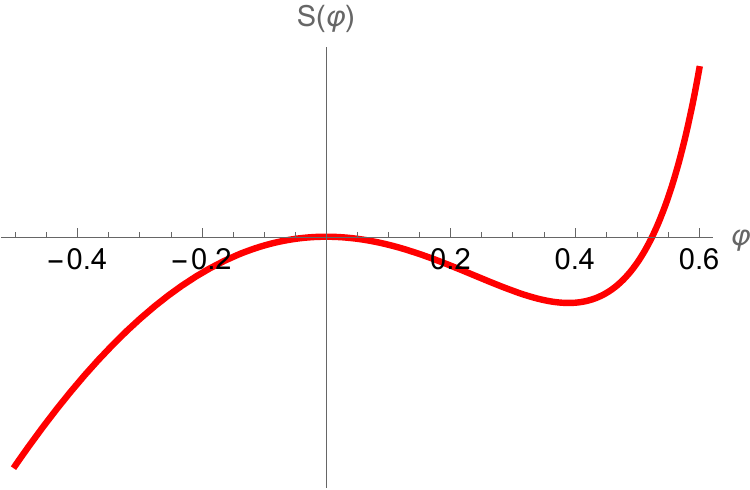} \;\;
\includegraphics[width=42mm]{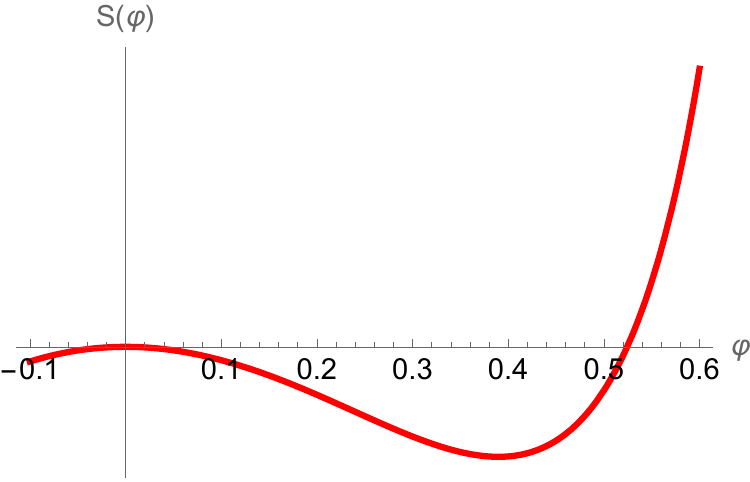} 
\caption{
Graph of the Sagdeev pseudopotential \eqref{sagdI} for $V = 1.2$ (left) 
and a zoom near the root $\varphi = 0.52438$ (right).
}
\label{SagdKdVd2}
\end{figure}

To apply SPA we again use $\zeta = x - V t$ to derive the cold ion density 
\beq
n_i = \frac{1}{\sqrt{1-\tfrac{2 \varphi}{V^2}}},
\label{densI}
\eeq
reminiscent of \eqref{dust} and use that to obtain the Sagdeev pseudopotential 
\beq
S(\varphi) = V^2 \left( 1 - \sqrt{1-\tfrac{2 \varphi}{V^2}} \right) 
    + \left( 1 - \mathrm{e}^\varphi \right).
\label{sagdI}
\eeq
From 
\beq 
S''(\varphi) = \frac{1}{V^2} \left( 1 - \tfrac{2 \varphi}{V^2} \right)^{-3/2} 
- \mathrm{e}^\varphi,
\label{sagdIz}
\eeq
one gets
\beq
S''(0) = \tfrac{1}{V^2} - 1.
\eeq
Therefore the ion-acoustic speed is $V_a = 1$.
There are no negative roots and there can only be one positive root before the 
limit $\varphi_{\mathrm{lim}} = \frac{V^2}{2}$ is reached.
The necessary condition is $S(\varphi_{\mathrm{lim}})\geqslant 0$.
Then 
\beq
S(\varphi_{\mathrm{lim}}) = V^2 + 1 - \mathrm{e}^{\frac{V^2}{2}} = 0
\label{sagdlimI}
\eeq 
yields $V = V_{\mathrm{lim}} = 1.5852$ and $\varphi_{\mathrm{lim}} = 1.25643$.
Hence, $1 < V < 1.5852$ is needed.
For each $V$ in that interval, $S(\varphi)=0$ then determines the value of the 
positive root.
A list of these roots (each corresponding to a value of $V$) is given 
in table~\ref{SagdKdVi}.
Figures~\ref{SagdKdVd1} and~\ref{SagdKdVd2} illustrate the shape of the Sagdeev 
pseudopotential \eqref{sagdI} for $V = 1.01$ and $V = 1.2$, respectively, 
together with zooms near the roots $\varphi = 0.02978$ and $\varphi = 0.52438$.
These roots are obtained by numerically solving $S(\varphi) = 0$. 

The actual graph of $\varphi(\zeta)$ can then be obtained by numerical 
integration of Poisson's equation \eqref{poisagain} for $S(\varphi)$ in 
\eqref{sagdI}, that is,     
\beq
\pd{^2\varphi}{\zeta^2} 
+ \frac{1}{\sqrt{1-\tfrac{2 \varphi}{V^2}}} - \mathrm{e}^{\varphi} = 0.
\label{poisIzeta}
\eeq 
 
\subsection*{Reductive perturbation theory}
Turning now to the reductive perturbation approach, 
we use the widely used  
(see e.g. \citet{Varghese-etal-physd-2025} and references therein) 
stretching
\beq
\xi = \eps^{1/2}(x - M t), \qquad \tau = \eps^{3/2} t,
\label{stretchI}
\eeq
and the expansions
\beqa 
n_i &=& 1 + \eps n_{i1} + \eps^2 n_{i2} + ...\ , \\
u_i &=& \eps u_{i1} + \eps^2 u_{i2} + ...\ , \\
\varphi &=& \eps \varphi_1 + \eps^2 \varphi_2 + ...\ , 
\eeqa
yielding the following results 
\beqa
\eps^{3/2}: 
&\qquad& M \pd{n_{i1}}{\xi} - \pd{u_{i1}}{\xi} = 0 ,  \nnn 
\eps^{5/2}: 
&\qquad& \pd{n_{i1}}{\tau} - M \pd{n_{i2}}{\xi} + \pd{u_{i2}}{\xi}
    + \pd{}{\xi}(n_{i1} u_{i1}) = 0 , \nnn
\eps^{3/2}: 
&\qquad& M \pd{u_{i1}}{\xi} - \pd{\varphi_1}{\xi} = 0 , \nnn
\eps^{5/2}: 
&\qquad& \pd{u_{i1}}{\tau} - M \pd{u_{i2}}{\xi} 
    + u_{i1} \pd{u_{i1}}{\xi} + \pd{\varphi_2}{\xi} = 0.
\label{interI}
\eeqa
Finally, \eqref{poisI} gives
\beq
\eps^2 \pd{^2 \varphi_1}{\xi^2} + (1 + \eps n_{i1} + \eps^2 n_{i2}) 
- (1 + \eps \varphi_1 + \eps^2 \varphi_2 + \tfrac{1}{2} \eps^2 \varphi_1^2) = 0.
\label{cpoisI}
\eeq 
Elimination of the terms at order $\eps^{3/2}$ in \eqref{interI} and order
$\eps$ in \eqref{cpoisI} yields
\beq
n_{i1} = \tfrac{1}{M^2} \varphi_1 = \varphi_1,
\eeq
and thus $M = 1$ in the stretching \eqref{stretchI}. 
So, $M$ matches the ion-acoustic speed (i.e. $V_a = 1$) 
established before.
The main difference is that in 
the SPA method $V$ represents 
the soliton speed with respect to the so-called laboratory frame.

Continuing with the higher-order terms in \eqref{interI} and \eqref{cpoisI},  
after elimination of $n_{i2}, u_{i2},$ and $\varphi_2$, leads to the 
well-known KdV equation, 
\beq
2 \pd{\varphi_1}{\tau} + 2 \varphi_1 \pd{\varphi_1}{\xi} 
    + \pd{^3\varphi_1}{\xi^3} = 0.
\label{kdvI}
\eeq
\subsection*{Comparison of the results from SPA and RPT}
To compare solutions obtained by SPA with solutions of \eqref{kdvI}, 
we move to a frame co-moving with the soliton with respect to the 
earlier stretching. 
Therefore, we set 
\beq
\zeta = \xi - v \tau,
\label{X}
\eeq
and
\beq
V = 1 + v, 
\label{W}
\eeq
which is the soliton velocity in the laboratory frame.
Hence, $v = V-1$ which will be used in the discussion below.
Using \eqref{X}, KdV equation \eqref{kdvI} is transformed into
\beq
- 2 v \od{\varphi_1}{\zeta} + 2 \varphi_1 \od{\varphi_1}{\zeta} 
  + \od{^3\varphi_1}{\zeta^3} = 0,
\label{kdvIagain}
\eeq
which has the well-known solution 
\beq
\varphi 
= 3 v \, \sech^2 \left( \sqrt{\tfrac{v}{2}} \zeta \right)
= 3 v \, \sech^2 \left( \sqrt{\tfrac{v}{2}} (\xi - v \tau) \right), 
\label{solitonI} 
\eeq
using \eqref{X}.
The maximum amplitude $3 v = 3 (V-1)$ of \eqref{solitonI} 
is reached at $\zeta=0$ and this amplitude increases linearly with $V$.
In SPA, the amplitude of the solitary wave is given by the value of the 
positive root of $S(\varphi)$, and the graph of $\varphi(\zeta)$ has to be 
obtained from a numerical integration of the Poisson equation \eqref{poisIzeta}.
The maximum amplitudes of the solitons computed with both methods for various 
choices of $V$ are given in table~\ref{SagdKdVi}.
The surprising conclusion is that the linearized equations seem to overestimate 
the solitary wave amplitude when the nonlinearities are fully included in the 
description.

There is a caveat: In the derivation of KdV solitons with RPT the 
nonlinearities are limited to second order. 
Thus for consistency, one can only allow perturbations of order 0.1 to 0.2.
Therefore, solutions \eqref{solitonI} with too large an amplitude might not 
reflect physical reality.
Although mathematically speaking, for large amplitudes KdV solitons can have 
interesting properties, the KdV equation and its solutions would then fail 
to accurately describe the physical model application.

\begin{table}
\begin{center}
\caption{Comparison of small solitary wave amplitudes computed with 
SPA and RPT.}  
\label{SagdKdVi} 
\begin{tabular}{ccccc}
${\mathbf{V}}$ & \textbf{Max.\ ampl.\ (SPA)} &  ${\mathbf{v}}$ & 
\textbf{Max.\ ampl.\ (KdV soliton)} & \textbf{Difference (KdV-SPA)}\\[3pt]
   1.01     & 0.02978           & 0.01 & 0.03          & 0.00022     \\
   1.02     & 0.05912           & 0.02 & 0.06          & 0.00088     \\
   1.03     & 0.08803           & 0.03 & 0.09          & 0.00197     \\
   1.04     & 0.11653           & 0.04 & 0.12          & 0.00347     \\
   1.05     & 0.14463           & 0.05 & 0.15          & 0.00537     \\
   1.06     & 0.17234           & 0.06 & 0.18          & 0.00766     \\
   1.07     & 0.19967           & 0.07 & 0.21          & 0.01033     \\
   1.08     & 0.22663           & 0.08 & 0.24          & 0.01337     \\
   1.09     & 0.25322           & 0.09 & 0.27          & 0.01678     \\
   1.10     & 0.27947           & 0.10 & 0.30          & 0.02053     \\
   1.20     & 0.52438           & 0.20 & 0.60          & 0.07662     \\
   1.30     & 0.74222           & 0.30 & 0.90          & 0.15778     \\
   1.40     & 0.93827           & 0.40 & 1.20          & 0.26173     \\
   1.50     & 1.11647           & 0.50 & 1.50          & 0.38353     \\[3pt]
\end{tabular}
\end{center}
\end{table}

One cannot know how reliable the KdV results are for a given model unless a 
comparison is made with either methods where the nonlinearities are kept in 
full or with physical experiments. 
Fortunately, for ion-acoustic waves in plasmas the Sagdeev pseudopotential 
method can be applied for a great many models. 
Nevertheless, quantitative comparisons have rarely been made.

\begin{figure}
\centering
\includegraphics[width=42mm]{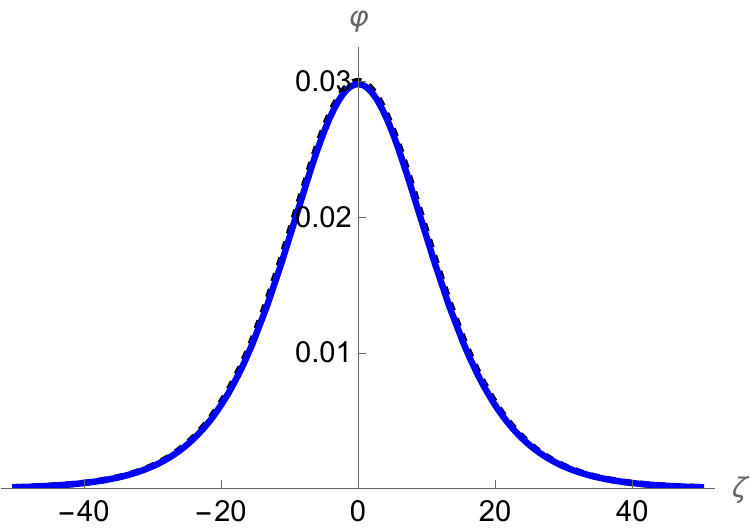} \;\;
\includegraphics[width=42mm]{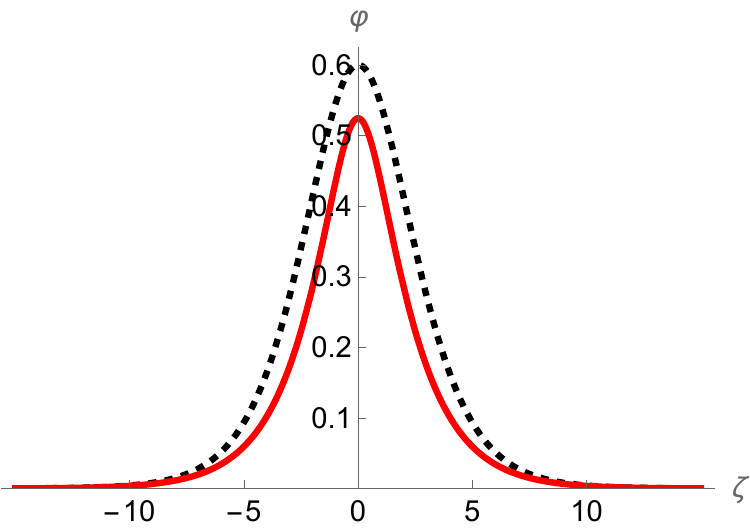}
\caption{
Comparison of the graphs of bright ion-acoustic solitons computed with SPA 
and RPT.
The solid curves come from numerical integration of Poisson's equation for 
$V = 1.01$ (left) and $V = 1.20$ (right).
The dashed curves show the $\sech$ squared profile in \eqref{solitonI} with 
$v = 0.01$ (left) and $v = 0.2$ (right).}
\label{IAKdVSagd}
\end{figure}

As seen in table~\ref{SagdKdVi} and figure~\ref{IAKdVSagd} where $v=V-1$, 
for very small amplitudes of $\varphi$ (computed with SPA) both curves coincide 
to a large extend, but when the maximum amplitude of $\varphi$ reaches 0.2 (and 
beyond) the KdV solutions tend to overestimate the fully nonlinear solutions. 
This is perhaps not what one would expect because the KdV description caps the 
nonlinearities at quadratic terms whereas SPA keeps the nonlinearities as they 
appear in the model equations.
Note also that as the amplitudes increase with the velocities (mostly not 
linearly), the solitons become narrower regardless of the method being used.  



\begin{thebibliography}{}

\bibitem[Ablowitz \& Clarkson(1991)]{AblowitzClarkson1991}
\textsc{Ablowitz}, M. J. \& \textsc{Clarkson}, P. A. 1991
\textit{Solitons, Nonlinear Evolution Equations and Inverse Scattering}.
London Mathematical Society Lecture Notes Series, vol.\ 149. 
Cambridge University Press.

\bibitem[Ablowitz \& Segur(1981)]{AblowitzSegur1981}
\textsc{Ablowitz}, M. J. \& \textsc{Segur}, H.  1981
\textit{Solitons and the Inverse Scattering}. 
SIAM Studies in Applied Mathematics, vol. 4. SIAM.

\bibitem[Bacha \& Tribeche(2013)]{BachaTribeche-jpp-2013}
\textsc{Bacha}, M. \& \textsc{Tribeche}, M.  2013
Nonlinear dust-ion acoustic waves in a dusty plasma with non-extensive 
electrons and ions. 
\textit{J.\ Plasma Phys.\/} \textbf{79}, 569--576.

\bibitem[Cairns \etal(1995)]{Cairns1995}
\textsc{Cairns}, R. A., \textsc{Mamun}, A. A., \textsc{Bingham}, R., 
\textsc{Bostr\"om}, R, \textsc{Dendy}, R. O.,
\textsc{Nairn}, C. M. C. \& \textsc{Shukla}, P. K. 1995
Electrostatic solitary structures in non-thermal plasmas.
\textit{Geophys.\ Res.\ Lett.\/} \textbf{22}, 2709--2712.

\bibitem[Drazin \& Johnson(1989)]{Drazin1989}
\textsc{Drazin}, P. G. \& \textsc{Johnson}, R. S.  1989
\textit{Solitons: An Introduction}. 
Cambridge Texts in Applied Mathematics. Cambridge University Press.

\bibitem[Gardner \etal(1967)]{Gardner1967}
\textsc{Gardner}, C. S., \textsc{Greene}, J. M., 
\textsc{Kruskal}, M. D. \& \textsc{Miura}, R. M.  1967
Method for solving the Korteweg-de Vries equation.
\textit{Phys.\ Rev.\ Lett.\/} \textbf{19}, 1095--1097.

\bibitem[Gardner \etal(1974)]{Gardner1974}
\textsc{Gardner}, C. S., \textsc{Greene}, J. M., \textsc{Kruskal}, M. D. 
\& \textsc{Miura}, R. M.  1974
Korteweg-de Vries equations and generalizations. VI. methods for exact solutions.
\textit{Commun.\ Pure Appl.\ Maths.\/} \textbf{27}, 97--133.

\bibitem[Gill, Kaur \& Saini (2005)]{Gilletal-jpp-2005}
\textsc{Gill}, T. S., \textsc{Kaur}, H. \& \textsc{Saini}, N. S.  2005.
A study of ion-acoustic solitons and double layers in a multispecies
collisionless weakly relativistic plasma. 
\textit{J.\ Plasma Phys.\/} \textbf{71}, 23--34.

\bibitem[Gesztesy, Schweiger \& Simon(1991)]{Gesztesy1991}
\textsc{Gesztesy}, F., \textsc{Schweiger}, W. \& \textsc{Simon}, B.  1991
Commutation methods applied to the mKdV-equation. 
\textit{Trans.\ Amer.\ Math.\ Soc.\/} \textbf{324}, 465--525.

\bibitem[Hereman \& G\"{o}kta\c{s}(2024a)]{HeremanGoktas-hirota-2024}
\textsc{Hereman}, W. \& \textsc{G\"{o}kta\c{s}}, \"{U}.  2024a 
Symbolic computation of solitary wave solutions and solitons through 
homogenization of degree.
\textit{Nonlinear and Modern Mathematical Physics}. 
Springer Proceedings in Mathematics \& Statistics 
(eds.\ S. Manukure and W.-X. Ma), vol.\ 459, pp.\ 101--164. Springer.  

\bibitem[Hereman \& G\"{o}kta\c{s}(2024b)]{HeremanGoktas-gardner-2024}
\textsc{Hereman}, W. \& \textsc{G\"okta\c{s}}, \"U.  2024b 
Using symmetries to investigate the complete integrability, solitary wave 
solutions and solitons of the Gardner equation.
\textit{Math.\ Computat.\ Applics.\/} \textbf{29}, 91. 

\bibitem[Hirota(1971)]{HirotaKdV1971}
\textsc{Hirota}, R.  1971
Exact solution of the Korteweg-de Vries equation for multiple collisions 
of solitons.
\textit{Phys.\ Rev.\ Lett.\/} \textbf{27}, 1192--1194.

\bibitem[Hirota(1972)]{HirotamKdV1972}
\textsc{Hirota}, R.  1972
Exact solution of the modified Korteweg-de Vries equation for multiple
collisions of solitons.
\textit{J.\ Phys.\ Soc.\ Japan} \textbf{33}, 1456--1458.

\bibitem[Hirota(2004)]{Hirota2004}
\textsc{Hirota}, R.  2004
\textit{The Direct Method in Soliton Theory}. 
Cambridge Texts in Mathematics, vol.\ 155. 
Cambridge University Press.

\bibitem[Kalita \& Das(2017)]{KalitaDas2017}
\textsc{Kalita}, B. C. \& \textsc{Das}, S.  2017
Comparative study of dust ion acoustic Korteweg–de Vries and modified 
Korteweg–de Vries solitons in dusty plasmas with variable temperatures.
\textit{J.\ Plasma Phys.\/} \textbf{83}, 905830502.

\bibitem[Korteweg \& de Vries(1895)]{KdV1895}
\textsc{Korteweg}, D. J. \& \textsc{de Vries}, G.  1895
XLI On the change of form of long waves advancing in a rectangular canal,
and on a new type of long stationary waves.
\textit{Lond.\ Edinb.\ Dublin Phil.\ Mag.\ J.\ Sci.\/} \textbf{39}, 422--443.

\bibitem[Miura, Gardner \& Kruskal(1968)]{Miura-etal-jmp-1968}
\textsc{Miura}, R. M., \textsc{Gardner}, C. S. 
\& \textsc{Kruskal}, M. D.  1968
Korteweg-de Vries equation and generalizations. 
II. Existence of conservation laws and constants of motion.
\textit{J.\ Math.\ Phys.\/} \textbf{9}, 1204--1209.

\bibitem[Nakamura \& Tsukabayashi(2009)]{Nakamura2009}
\textsc{Nakamura}, Y. \& \textsc{Tsukabayashi}, I.  2009
Modified Korteweg-de Vries ion-acoustic solitons in a plasma.
\textit{J.\ Plasma Phys.\/} \textbf{34}, 401--415.

\bibitem[Nasipuri \etal(2025)]{Nasipuri-etal-ijp-2025}
\textsc{Nasipuri}, S., \textsc{Chandra}, S., \textsc{Ghosh}, U. N., 
\textsc{Das}, C. \& \textsc{Chatterjee}, P.  2025  
Study of breather structures in the framework of Gardner equation in 
electron-positron-ion plasma.
\textit{Indian J.\ Phys.\/}, doi: 10.1007/s12648-025-03594-0, in press. 

\bibitem[Olivier \& Verheest(2020)]{OlivierVerheest-pp-2020} 
\textsc{Olivier}, C. P. \& \textsc{Verheest}, F.  2020.
Overtaking collisions of double layers and solitons: 
Tripolar structures and dynamical polarity switches.
\textit{Phys.\ Plasmas} \textbf{27}, 062303. 

\bibitem[Olivier \& Verheest(2022)]{OlivierVerheest-jpp-2022} 
\textsc{Olivier}, C. P. \& \textsc{Verheest}, F.  2022.
A new reductive perturbation formalism for ion acoustic cnoidal waves.
\textit{J.\ Plasma Phys.\/} \textbf{88}, 905880601. 

\bibitem[Olivier, Verheest \& Maharaj(2016)]{Olivier-etal-jpp-2016} 
\textsc{Olivier}, C. P., \textsc{Verheest}, F. 
\& \textsc{Maharaj}, S. K.  2016.
A small-amplitude study of solitons near critical plasma compositions.
\textit{J.\ Plasma Phys.\/} \textbf{82}, 905820605.

\bibitem[Remoissenet(1999)]{Remoissenet1999}
\textsc{Remoissenet}, M.  1999
\textit{Waves Called Solitons: Concepts and Experiments} 
3rd edn.\ Advanced Texts in Physics. Springer.

\bibitem[Sagdeev(1966)]{Sagdeev1966}
\textsc{Sagdeev}, R. Z.  1966
Cooperative phenomena and shock waves in collisionless plasmas.
In \textit{Reviews of Plasma Physics} (ed.\ M.\ A.\ Leontovich), vol.\ 4, 
pp.\ 23–-91. Consultants Bureau.
%

\bibitem[Scott Russell(1844)]{ScottRussell1844}
\textsc{Scott Russell}, J.  1844
Report on Waves. In 
\textit{Report 14th Meeting British Association Advancement of
Science}, pp.\ 311-–390. John Murray.
%

\bibitem[Singh \& Kourakis(2025)]{SinghKourakis2025}
\textsc{Singh}, K. \& \textsc{Kourakis}, I.  2025
Generalized analytical solutions of a Korteweg–de Vries (KdV)
equation with arbitrary real coefficients: Association with the
plasma-fluid framework and physical interpretation. 
\textit{Wave Motion} \textbf{132}, 103443.

\bibitem[Torven(1981)]{Torven-prl-1981} 
\textsc{Torv\'en}, S.  1981
Modified Korteweg-de Vries equation for propagating double layers in plasmas. 
\textit{Phys.\ Rev.\ Lett.\/} \textbf{47}, 1053--1056. 

\bibitem[Varghese \etal(2025)]{Varghese-etal-physd-2025}
\textsc{Varghese}, S. S., \textsc{Singh}, K., \textsc{Verheest}, F. 
\& \textsc{Kourakis}, I.  2025 
Integrable nonlinear PDEs as evolution equations derived from multi-ion 
fluid plasma models. 
\textit{Physica D} \textbf{472}, 134527.

\bibitem[Verheest(2000)]{VerheestASSL2000}
\textsc{Verheest}, F.  2000
\textit{Waves in Dusty Space Plasmas}. 
Astrophysics and Space Science Library, vol.\ 245. Kluwer.

\bibitem[Verheest, Hellberg \& Olivier(2020)]{Verheest-etal-pp-2020}
\textsc{Verheest}, F., \textsc{Hellberg}, M. A. 
\& \textsc{Olivier}, C. P.  2020
Electrostatic flat-top solitons near double layers and triple root structures
in multispecies plasmas: How realistic are they? 
\textit{Phys.\ Plasmas} \textbf{27}, 062306.

\bibitem[Verheest \& Hereman(2019)]{VerheestHereman-jpp-2019}
\textsc{Verheest}, F. \& \textsc{Hereman}, W. A.  2019.
Collision of acoustic solitons and their electric fields in plasmas at
critical compositions. 
\textit{J.\ Plasma Phys.\/} \textbf{85}, 905850106.

\bibitem[Verheest \& Olivier(2024)]{CnoidalSadha2024}
\textsc{Verheest}, F. \& \textsc{Olivier}, C. P.  2024
Periodic nonlinear dust-acoustic waves in multispecies dusty plasmas.
\textit{Phys.\ Plasmas} \textbf{31}, 083701.

\bibitem[Verheest \& Pillay(2008a)]{SadhaNT12008a}
\textsc{Verheest}, F. \& \textsc{Pillay}, S. R.  2008a
Large amplitude dust-acoustic solitary waves and double layers in nonthermal
plasmas.
\textit{Phys.\ Plasmas} \textbf{15}, 013703.

\bibitem[Verheest \& Pillay(2008b)]{SadhaNT22008b} 
\textsc{Verheest}, F. \& \textsc{Pillay}, S. R.  2008b
Dust-acoustic solitary structures in plasmas with nonthermal electrons and 
positive dust.
\textit{Nonlinear Process.\ Geophys.\/} \textbf{15}, 551--555. 

\bibitem[Verheest, Olivier \& Hereman(2016)]{Verheestetal-jpp-2016}
\textsc{Verheest}, F., \textsc{Olivier}, C. P. \& \textsc{Hereman}, W. A. 2016.
Modified Korteweg-de Vries solitons at supercritical densities 
in two-electron temperature plasmas.
\textit{J.\ Plasma Phys.\/} \textbf{82}, 905820208.

\bibitem[Wadati(1973)]{Wadati-jpsjpn-1973}
\textsc{Wadati}, M.  1973
The modified Korteweg-de Vries equation.
\textit{J.\ Phys.\ Soc.\ Japan} \textbf{34}, 1289--1296.

\bibitem[Washimi \& Taniuti(1966)]{Washimi1966}
\textsc{Washimi}, H. \& \textsc{Taniuti}, T.  1966
Propagation of ion-acoustic solitary waves of small amplitude.
\textit{Phys.\ Rev.\ Lett.\/} \textbf{17}, 996--997.

\bibitem[Watanabe(1984)]{Watanabe1984}
\textsc{Watanabe}, S.  1984
Ion acoustic soliton in plasma with negative ions.
\textit{J.\ Phys.\ Soc.\ Japan} \textbf{53}, 950--956.

\bibitem[Xie \& He(1999)]{XieHe-pp-1999}
\textsc{Xie}, B. \& \textsc{He}, K.  1999.
Dust-acoustic solitary waves and double layers in a dusty plasma with 
variable dus charge and two-temperature ions. 
\textit{Phys.\ Plasmas} \textbf{6}, 3808--3816.

\bibitem[Zabusky \& Kruskal(1965)]{Zabusky1965}
\textsc{Zabusky}, N. J. \& \textsc{Kruskal}, M. D.  1965
Interactions of solitons in a collsionless plasma and the recurrence of initial
states.
\textit{Phys.\ Rev.\ Lett.\/} \textbf{15}, 240--243.

\bibitem[Zhang \etal(2014)]{Zhang-etal-rmp-2014}
\textsc{Zhang}, D.-J., \textsc{Zhao}, S.-L., \textsc{Sun}, Y.-Y. 
\& \textsc{Zhou}, J. 2014.
Solutions to the modified Korteweg–de Vries equation.
\textit{Rev.\ Math.\ Phys.\/} \textbf{26}, 1430006.

\end{thebibliography}
\end{document}